\input lanlmac
\input epsf.tex
\input amssym.tex
\font\sc=cmcsc10
\overfullrule=0pt
\def\figbox#1#2{\epsfxsize=#1%
\vcenter{\hbox{\epsfbox{#2}}}}
\def\raisedfigbox#1#2#3{\epsfxsize=#1%
\raise#3\hbox{\epsfbox{#2}}}
\newcount\figno
\figno=0
\def\fig#1#2#3{
\par\begingroup\parindent=0pt\leftskip=1cm\rightskip=1cm\parindent=0pt
\baselineskip=11pt
\global\advance\figno by 1
\midinsert
\epsfxsize=#3
\centerline{\epsfbox{#2}}
\vskip 12pt
{\bf Fig.\ \the\figno:} #1\par
\endinsert\endgroup\par
}
\def\figlabel#1{\xdef#1{\the\figno}%
\writedef{#1\leftbracket \the\figno}%
}

\def\e#1{{\rm e}^{#1}}
\def\pre#1{{\tt
#1}}
\def\d{{\rm d}}

\def\Rc{{\check R}}

\newcount\propno
\propno=1
\def\prop#1#2\par{\xdef#1{\the\propno}%
\medbreak\noindent{\sc Proposition \the\propno.\enspace}{\sl #2}\medskip%
\global\advance\propno by1}
\def\thmn#1#2\par{
\medbreak\noindent{\sc Theorem #1.\enspace}{\sl #2\par}\medskip%
}
\newcount\thmno
\thmno=1
\def\thm#1#2\par{\xdef#1{\the\thmno}%
\thmn{\the\thmno}{#2}%
\global\advance\thmno by1}
\newcount\conjno
\conjno=1
\def\conj#1#2\par{\xdef#1{\the\conjno}%
\medbreak\noindent{\sc Conjecture \the\conjno.\enspace}{\sl #2}\medskip%
\global\advance\conjno by1}
\newcount\lemno
\lemno=1
\def\lemma#1#2\par{\xdef#1{\the\lemno}%
\medbreak\noindent{\sc Lemma \the\lemno.\enspace}{\sl #2}\medbreak%
\global\advance\lemno by1}
\def\corol#1\par{%
\medbreak\noindent{\sc Corollary.\enspace}{\sl #1}\medskip}
\long\def\example#1\par{%
\medbreak\noindent{\sc Example.\enspace}#1\medskip}
\def\qed{\nobreak\hfill\vbox{\hrule height.4pt%
\hbox{\vrule width.4pt height3pt \kern3pt\vrule width.4pt}\hrule height.4pt}\medskip\goodbreak}
\lref\RS{A.V. Razumov and Yu.G. Stroganov, 
{\sl Combinatorial nature of ground state vector of $O(1)$ loop model},
{\it Theor. Math. Phys.} 
{\bf 138} (2004) 333--337; {\it Teor. Mat. Fiz.} 138 (2004) 395--400, \pre{math.CO/0104216}.}
\lref\BdGN{M.T. Batchelor, J. de Gier and B. Nienhuis,
{\sl The quantum symmetric XXZ chain at $\Delta=-1/2$, alternating sign matrices and 
plane partitions},
{\it J. Phys.} A34 (2001) L265--L270,
\pre{cond-mat/0101385}.}
\lref\dG{J. de Gier, {\sl Loops, matchings and alternating-sign matrices},
{\it Discr. Math.} 298 (2005), 365--388,
\pre{math.CO/0211285}.}
\lref\Rob{D.P.~Robbins, {\sl Symmetry classes of alternating-sign matrices},
\pre{math.CO/0008045}.}
\lref\PRdG{P. A. Pearce, V. Rittenberg and J. de Gier, 
{\sl Critical Q=1 Potts Model and Temperley--Lieb Stochastic Processes},
\pre{cond-mat/0108051}.}
\lref\RSb{A.V. Razumov and Yu.G. Stroganov, 
{\sl $O(1)$ loop model with different boundary conditions and symmetry classes of alternating-sign matrices},
{\it Theor. Math. Phys.} 
{\bf 142} (2005) 237--243; {\it Teor. Mat. Fiz.} 142 (2005) 284--292,
\pre{cond-mat/0108103}.}
\lref\RSc{A.V. Razumov and Yu.G. Stroganov, 
{\sl Bethe roots and refined enumeration of alternating-sign matrices},
{\it J. Stat. Mech. (2006) P07004},
\pre{math-ph/0605004}.}
\lref\RSZJ{A.V. Razumov, Yu.G. Stroganov and P. Zinn-Justin,
work in progress.}
\lref\PRdGN{P. A. Pearce, V. Rittenberg, J. de Gier and B. Nienhuis,
{\sl Temperley--Lieb Stochastic Processes}, 
{\it J. Phys. A} {\bf 35 } (2002) L661-L668, \pre{math-ph/0209017}.}
\lref\dGR{J.~de~Gier and V.~Rittenberg,
{\sl Refined Razumov--Stroganov conjectures for open boundaries},
{\it JSTAT} (2004) P09009,
\pre{math-ph/0408042}.}
\lref\MNosc{S. Mitra and B. Nienhuis, {\sl 
Osculating random walks on cylinders}, in
{\it Discrete random walks}, 
DRW'03, C. Banderier and
C. Krattenthaler edrs, Discrete Mathematics and Computer Science
Proceedings AC (2003) 259-264, \pre{math-ph/0312036}.} 
\lref\Kup{G. Kuperberg, {\sl Symmetry classes of alternating-sign matrices under one roof},
{\it Ann. of Math.} (2) 156 (2002), no. 3, 835--866,
\pre{math.CO/0008184}.}
\lref\MNdGB{S. Mitra, B. Nienhuis, J. de Gier and M.T. Batchelor,
{\sl Exact expressions for correlations in the ground state 
of the dense $O(1)$ loop model}, 
{\it JSTAT} (2004) P09010,
\pre{cond-mat/0401245}.}
\lref\DF{P.~Di~Francesco, {\sl 
 A refined Razumov--Stroganov conjecture} I: 
     J. Stat. Mech. (2004) P08009, \pre{cond-mat/0407477}; II: 
     J. Stat. Mech. (2004) P11004, \pre{cond-mat/0409576}.}
\lref\DFb{P.~Di~Francesco,
{\sl Totally Symmetric Self-Complementary Plane Partitions and 
Quantum Knizhnik-Zamolodchikov equation: a conjecture},
J. Stat. Mech. (2006) P09008, \pre{cond-mat/0607499}.}
\lref\KP{M.~Kasatani and V.~Pasquier,
{\sl On polynomials interpolating between the stationary state of a O(n) model and a Q.H.E. ground state},
\pre{cond-mat/0608160}.}
\lref\LGV{B. Lindstr\"om, {\it On the vector representations of
induced matroids}, Bull. London Math. Soc. {\bf 5} (1973)
85--90\semi
I. M. Gessel and X. Viennot, {\it Binomial determinants, paths and
hook formulae}, Adv. Math. { \bf 58} (1985) 300--321. }
\lref\DFZJ{P.~Di Francesco and P.~Zinn-Justin, {\sl Around the Razumov--Stroganov conjecture:
proof of a multi-parameter sum rule}, {\it E. J. Combi.} 12 (1) (2005), R6,
\pre{math-ph/0410061}.}
\lref\Pas{V.~Pasquier, {\sl Quantum incompressibility and Razumov Stroganov type conjectures},
{\it Ann. Henri Poincar\'e} 7 (2006), 397--421,
\pre{cond-mat/0506075}.}
\lref\FR{I.B.~Frenkel and N.~Reshetikhin, {\sl Quantum affine Algebras and Holonomic Difference Equations},
{\it Commun. Math. Phys.} 146 (1992), 1--60.}
\lref\DFZJb{P. Di Francesco and P. Zinn-Justin,
{\sl Inhomogeneous model of crossing loops and multidegrees of some algebraic varieties},
{\it Commun. Math. Phys.} 262 (2006), 459--487,
\pre{math-ph/0412031}.}
\lref\KZJ{A. Knutson and P. Zinn-Justin,
{\sl A scheme related to Brauer loops},
to appear in {\it Advances In Mathematics},
\pre{math.AG/0503224}.}
\lref\DFZJc{P.~Di Francesco and P.~Zinn-Justin, 
{\sl Quantum Knizhnik--Zamolodchikov equation, generalized Razumov--Stroganov sum rules 
and extended Joseph polynomials}, 
{\it J. Phys. A} 38 (2005) L815--L822, \pre{math-ph/0508059}.}
\lref\DFZJd{P.~Di Francesco and P.~Zinn-Justin, {\sl From Orbital Varieties to Alternating 
Sign Matrices}, extended abstract for FPSAC'06 (2006), \pre{math-ph/0512047}.}
\lref\DFZJZ{P.~Di~Francesco, P.~Zinn-Justin and J.-B.~Zuber,
{\sl A Bijection between classes of Fully Packed Loops and Plane Partitions},
{\it E. J. Combi.} 11(1) (2004), R64,
\pre{math.CO/0311220}.}
\lref\DFZ{P.~Di~Francesco and J.-B.~Zuber, {\sl On FPL 
configurations with four sets of nested arches}, 
{\it JSTAT} (2004) P06005, \pre{cond-mat/0403268}.}
\lref\DFZJZb{P.~Di~Francesco, P.~Zinn-Justin and J.-B.~Zuber,
{\sl Determinant Formulae for some Tiling Problems and Application to Fully Packed Loops}, 
{\it Annales de l'Institut Fourier} 55 (6) (2005), 2025--2050, \pre{math-ph/0410002}.} 
\lref\Oka{S. Okada, 
{\sl  Enumeration of Symmetry Classes of Alternating Sign Matrices and Characters of Classical Groups}, 
{\it J. Algebr. Comb.} 23 (2006), 43--69,
\pre{math.CO/0408234}.}
\lref\IZERKO{A. Izergin, {\sl Partition function of the six-vertex
model in a finite volume}, {\it Sov. Phys. Dokl.} {\bf 32} (1987) 878-879;
V. Korepin, {\sl Calculation of norms of Bethe wave functions},
{\it Comm. Math. Phys.} {\bf 86} (1982) 391-418.}
\lref\Kratt{F. Caselli and C.~Krattenthaler, {\sl Proof of two conjectures of
Zuber on fully packed loop configurations}, {\it J. Combin. Theory
Ser.} {\bf A 108} (2004), 123--146, \pre{math.CO/0312217}.  }
\lref\DF{P.~Di Francesco, {\sl 
 Inhomogeneous loop models with open boundaries }, 
{\it J. Phys. A: Math. Gen.} {\bf 38} 6091 (2005), \pre{math-ph/0504032}\semi
{\sl Boundary $q$KZ equation and generalized Razumov--Stroganov sum rules for open IRF models}, 
{\it J. Stat. Mech.} P11003 (2005), \pre{math-ph/0509011}.}
\lref\CZJ{L.~Cantini and P.~Zinn-Justin, work in progres.}
\lref\Bre{D. Bressoud, {\sl Proofs and Confirmations: The Story of
the Alternating Sign Matrix Conjecture}, Cambridge Univ. Pr., 1999.}
\lref\JM{M.~Jimbo and T.~Miwa, {\sl Algebraic Analysis of Solvable Lattice Models},
CBMS 85, American Mathematical Society (1995).}
\lref\Smi{F.A.~Smirnov,
{\sl A general formula for soliton form factors in the quantum sine-Gordon model},
{\it J. Phys.} A 19 (1986), L575--L578.}
\lref\ROB{D. Robbins, {\it The story of 1,2,7,42,429,7436,...}, 
{\it Mathl. Intelligencer} {\bf 13} No.2 (1991) 12-19.}
\lref\DFop{P.~Di~Francesco,
{\sl Open boundary Quantum Knizhnik-Zamolodchikov equation and the weighted 
enumeration of Plane Partitions with symmetries}, {\it J. Stat. Mech.} (2007) P01024,
\pre{math-ph/0611012}.}
\lref\Zei{D.~Zeilberger,
{\sl Proof of the alternating sign matrix conjecture},
{\it Elec. J. Comb.} 3(2) (1996), R13, \pre{math.CO/9407211}.}
\lref\PZRS{P. Zinn-Justin, {\sl Proof of Razumov-Stroganov conjecture for some 
infinite families of link patterns}, \pre{math.CO/0607183}.}
\lref\PZspin{P. Zinn-Justin, {\sl Combinatorial point for higher spin loop models}, 
\pre{math-ph/0603018}.}
\Title{}
{
\vbox{\centerline{Quantum Knizhnik--Zamolodchikov equation,}
\medskip
\centerline{Totally Symmetric Self-Complementary Plane Partitions}
\medskip
\centerline{and Alternating Sign Matrices}}
}
\bigskip\bigskip
\centerline{P. Zinn-Justin \footnote{${}^\star$}
{Laboratoire de Physique Th\'eorique et Mod\`eles Statistiques
(CNRS, UMR 8626); Univ Paris-Sud, Orsay, F-91405.}}
\bigskip
\centerline{P. Di Francesco\footnote{${}^\diamond$}
{Service de Physique Th\'eorique de Saclay,
CEA/DSM/SPhT, URA 2306 du CNRS, 
F-91191 Gif sur Yvette Cedex.}}
\vskip0.5cm
\noindent
We present multiresidue integral formulae for
partial sums in the basis of link patterns of the polynomial solution to the level $1$ 
$U_q(\widehat{\goth sl_2})$ quantum Knizhnik--Zamolodchikov
equation at generic values of the quantum parameter $q$. These
allow for rewriting and generalizing a recent conjecture [Di Francesco '06] connecting
the above to generating polynomials for weighted Totally Symmetric
Self-Complementary Plane Partitions. We reduce the corresponding
conjectures to a single integral identity.

\bigskip

\def\Rc{{\check R}}
\def\LP{{\cal L}}

\Date{03/2007}
%
%
%
%
\newsec{Introduction}
In the past few years, we have witnessed the development of
an ever increasingly complex set of
interrelations betwen various combinatorial objects and some
integrable models: on the one hand, Alternating Sign Matrices, Fully
Packed Loops and Plane Partitions; on the other hand, XXZ and
related spin chains, and lattice loop models. These will be reviewed below,
the goal of this paper being to
provide some tools to prove the combinatorial conjectures that were
formulated, and in particular the conjecture of \DFb. We believe that the tools developed
here may apply to a number of other situations as well, for instance to the case of
different boundary conditions for the loop model (e.g. open boundary case, see 
conjectures in \DFop).

\subsec{Integrable loop model and the Razumov--Stroganov conjecture}
The Temperley--Lieb model of loops is defined on a semi-infinite cylinder of square lattice,
with even perimeter $2n$ whose edge centers are labelled $1,2,\ldots,2n$ 
counterclockwise. The configurations of the model are obtained by picking any of the two
possible face configurations 
$\vcenter{\hbox{\epsfbox{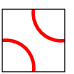}}}$ 
or 
$\vcenter{\hbox{\epsfbox{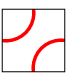}}}$ 
at
each face of the lattice. 
We see that the configurations of the model form either closed
loops or open curves joining boundary points by pairs, without any intersection between curves.
In fact, each configuration realizes a planar pairing of the boundary points via a
link pattern, namely a diagram in which $2n$ labelled and regularly spaced points
of a circle are connected by pairs via non-intersecting straight segments.
The set of link patterns over $2n$ points is denoted by $\LP_{2n}$, and its cardinality is 
$c_n=(2n)!/(n!(n+1)!)$. We also view $\pi\in \LP_{2n}$ as an
involution of ${\cal S}_{2n}$ without fixed points.

We moreover consider an inhomogeneous model in which probabilities are associated
to these face configurations depending on the row $i$ at which they
are inserted. We use the following parameterization of the
probabilities in terms of variables $z_i$:
\eqn\transmat{ T_n(t;z_1,\ldots,z_{2n})=\prod_{i=1}^{2n} \big({q\,z_i-q^{-1}t\over q\,t-q^{-1}z_i} 
\vcenter{\hbox{\epsfbox{mov1.eps}}}
+{z_i-t\over q\,t-q^{-1}z_i}
\vcenter{\hbox{\epsfbox{mov2.eps}}}
\big)}
where $T$ is the {\it transfer
matrix}\/ that adds one row to the semi-infinite cylinder, and here
$q=\e{2i\pi/3}$. $t$ is an additional variable which can be absorbed
into the $z_i$; however it is useful to introduce it because of the
commutation relations (due to integrability) $[T_n(t),T_n(t')]=0$ at $z_i$ fixed.

We may now ask what is the probability $P_\pi(z_1,\ldots,z_{2n})$ in random configurations of 
the model that the boundary points be pair-connected according to a given link pattern 
$\pi\in \LP_{2n}$. Forming the vector 
$P_n(z_1,\ldots,z_{2n})$ whose components are the
$P_\pi(z_1,\ldots,z_{2n})$ in a vector space with canonical
basis indexed by link patterns $\pi$,
we immediately see that
it satisfies the eigenvector condition
\eqn\evectP{ T_n(t;z_1,\ldots,z_{2n}) P_n(z_1,\ldots,z_{2n})=P_n(z_1,\ldots,z_{2n}) }
Eq.~\evectP\ does not specify the normalization of $P_n$, which is
given by the fact that the total probability is $1$. It is however
more convenient to consider another normalization of the solution of
Eq.~\evectP, which we denote by $\Psi_n$, and
such that its components $\Psi_\pi$ are {\it coprime}\/ polynomials in
the variables $z_i$. This only leaves a numerical multiplicative
constant to be fixed later.

The main 
conjectures concern the homogeneous limit
$z_1=\cdots=z_{2n}=1$. 
In \BdGN, it was noticed that: 
\item{(i)} the $\Psi_\pi/\Psi_{\pi_0}$ are
positive integers, where $\Psi_{\pi_0}$ is the smallest
component;
\item{(ii)} $\Psi_{max}/\Psi_{\pi_0}=A_{n-1}$, where $\Psi_{max}$ is
the largest component, and 
\item{(iii)} $\sum_{\pi\in \LP_{2n}}
\Psi_\pi/\Psi_{\pi_0}=A_n$, 

\noindent where $A_n$ is the number of Alternating
Sign Matrices (ASMs) of size $n$ (see the book by Bressoud \Bre\ for a complete saga
and references):
\eqn\An{
A_n={1!4!7!\cdots(3n-2)!\over n!(n+1)!(n+2)!\cdots(2n-1)!}
}
$A_n$ is also the number of Totally Symmetric Self-Complementary Plane
Partitions (TSSCPPs), which may be viewed as the tiling configurations
of a regular hexagon of edge size $2n$ drawn on the triangular lattice,
by means of elementary rhombi, and enjoying all possible symmetries 
of the hexagon.
Until recently, this fact did not seem particularly
relevant to this model, especially in view of the
Razumov--Stroganov (RS) conjecture \RS. The latter
claims that $\Psi_\pi/\Psi_{\pi_0}$ is the
number of Fully Packed Loops configurations (FPL) with connectivity
$\pi$. The latter are configurations of loops drawn on the edges of 
a square grid of size $n\times n$, such that at each vertex exactly two
edges are occupied by loop edges, and with the boundary condition that
every second edge of the boundary is occupied. Labelling the latter
$1,2,\ldots,2n$ allows for keeping track of pairings of boundary edges
via curves, hence to associate to each link pattern $\pi$ a set of FPL
configurations. 
Due to their definition, the FPLs are in simple bijection with the configurations
of the six-vertex model with domain-wall boundary conditions, as well as with ASMs.
The RS conjecture therefore gives an enumerative
interpretation of points (i) and (iii) (and in principle also of (ii)).
Note that on the
contrary, no bijection is known betteen ASMs and TSSCPPs.

Point (iii) was proved in \DFZJ\ via its generalization to the
inhomogeneous model, which reads $\sum_\pi \Psi_\pi =
Z_n(z_1,\ldots,z_{2n})$, where $Z_n$ is the Izergin--Korepin partition
function of the six-vertex model with domain-wall boundary conditions \IZERKO, 
which already appeared in \Kup\ in
the context of ASM enumeration. 

\subsec{TSSCPP conjecture and the minimal polynomial solutions to the $q$KZ equation}
It was noted in \DFZJ\ that Eq.~\evectP\ may be solved by writing instead the consequences
on $\Psi_n$ of the intertwining relations for the transfer matrix \transmat\
of the inhomogeneous loop model. 
The latter could eventually be reexpressed as the level 1 $U_q(\widehat{\goth sl_2})$ quantum Knizhnik--Zamolodchikov
($q$KZ) equation \FR\ for the groundstate vector $\Psi_n$, at $q=-e^{i\pi/3}$. When written in
components, this equation allows for expressing all polynomial components $\Psi_\pi$
in terms of $\Psi_{\pi_0}$, in a triangular manner. $\Psi_{\pi_0}$ is then determined
as the minimal degree polynomial subject to all divisibility properties inherited from 
the intertwining relations, and has a nice factorized form. The $q$KZ equation satisfied
by $\Psi_n$ may also alternatively be interpreted by stating that the components $\Psi_\pi$
form a polynomial representation of the affine Temperley-Lieb algebra, including
a weight $\tau=-q-q^{-1}=1$ per loop.

There is a natural $q$-deformation of the problem obtained either by looking for polynomial
representations of the affine Temperley-Lieb algebra with weight $\tau=-q-q^{-1}$ per
loop \Pas\ or equivalently in the form of the
polynomial solutions of the generic $q$ level 1 $U_q(\widehat{\goth sl_2})$ $q$KZ equation \DFZJc. It was noted
in \DFZJc\ and \KP\ that the properly normalized vector $\Psi_n$ displays nice combinatorial
properties in the homogeneous limit, but now as a function of $\tau$. It was indeed noted
that each ratio $\Psi_\pi/\Psi_{\pi_0}$, when expressed at $z_i=1$, appears to be 
a polynomial of $\tau$ with non-negative integer coefficients. It would of course have been
tempting to infer the existence of a $q$-deformed RS-type conjecture, in which 
these quantities could be interpreted as weighted sums of FPL configurations with fixed
connectivities. 

A first step was made in \DFb\ in this direction, by proposing a conjecture
for the sum of all components as a polynomial of $\tau$.
But surprisingly,
the right combinatorial object conjecturally underlying this sum is not
the FPLs or ASMs, but rather the TSSCPPs.\foot{Actually, some earlier works \DF\ had already 
unearthed some connections between the loop model in slightly deformed cylinder geometries
and refined TSSCPP counting, but at $\tau=1$.}

To understand the conjecture, it is
first necessary to express the TSSCPPs as sets of non-intersecting lattice paths (NILPs)
by noting that tiles in a fundamental domain of the hexagon may be organized into sets of 
non-intersecting broken lines propagating
within that region. Once this is done, the TSSCPPs may be viewed as configurations
of lattice paths which may have two possible types of steps, say vertical and diagonal
when represented on a square lattice.

The TSSCPP conjecture states that
\item{(iv)} $\sum_\pi \Psi_\pi/\Psi_{\pi_0}$ is the generating polynomial
for TSSCPPs of size $2n$ with a weight $\tau$ per vertical step in their NILP formulation.\par

\noindent Note that point (iv) reduces to point (iii) at $\tau=1$. 
The main purpose of this paper is to address the TSSCPP conjecture (iv).

It is interesting to note that refinements of TSSCPPs have been already related to
refined enumerations of ASMs (c.f. \ROB), but these concern always ``boundary''
weightings of configurations (e.g. weighting only the {\it last}\/ or {\it first}\/ 
step of the TSSCPPs, versus keeping track of the positions of $1$'s in the first 
row or column of the ASMs). Here we rather have a ``bulk'' weighting, that does not
distinguish the boundary steps. The question of finding a good interpretation
for the $\tau$ weighting in ASMs or FPLs remains open, and might shed some light
on a possible ASM--TSSCPP bijection.

\subsec{Plan}
In this paper, we first address the weighted enumeration of TSSCPPs (Section 2). 
The results are derived by use of the
formulation of the latter in terms of non-intersecting lattice paths (NILPs), and the
celebrated Lindstr\"om--Gessel--Viennot (LGV) determinantal formula \LGV, and take the form
of multiple contour (residue) integrals.

In a second step (Section 3), we shall write the space of polynomial
solutions of the $q$KZ equation in terms of multiple contour
(residue) integrals, by using a new set of vectors indexed by
sequences of integers (distinct from the so-called spin basis used in the 
language of the corresponding spin chains). Upon going to the
loop (link pattern) basis, 
this will give us integral formulae for {\it partial sums}\/ of components
of $\Psi$. In particular, we will get integral formulae for the maximal component
$\Psi_{max}$ as well as for the sum of all components $\sum_\pi \Psi_\pi$. The details of the
change of basis from this new set of vectors to link patterns is given
in appendix A, while the relation to spin basis components is detailed in appendix B.

We then show in Section 4 how the TSSCPP conjecture (iv) as well as (ii) and (iii) above are 
all consequences of a simple multiresidue integral
identity, whose proof is left to future work.

\newsec{TSSCPP}
\subsec{Definitions}
\fig{From TSSCPP to NILP. Including a weight $t_i$ per vertical step in the
$i$-th horizontal slice from top, the NILP configuration receives the weight
$t_2^2t_3t_4$.}{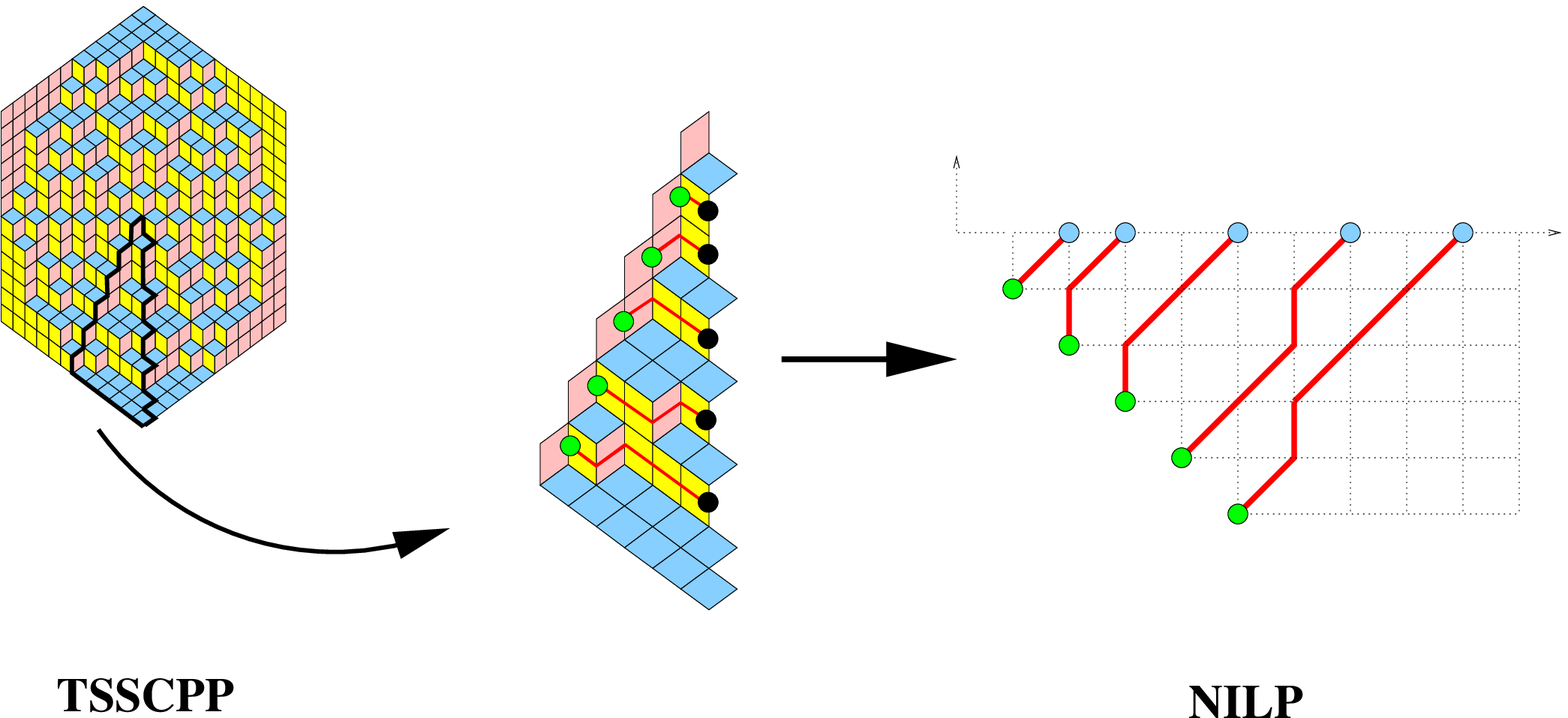}{13.cm}
\figlabel\tsscppnilp

Totally Symmetric Self-Complementary Plane Partitions are rhombus tilings of a 
regular hexagon of the triangular lattice, of size $2n\times 2n\times 2n$, enjoying 
rotational and reflection symmetries of the hexagon, implementing the self-complementation property 
that, when viewed as the $(1,1,1)$ direction perspective view of a pile of elementary 
cubes in the positive quadrant of ${\Bbb Z}^3$, the complement to the pile within the cube 
of size $2n$ is an identical copy of it. 

This allows to restrict TSSCPPs to a fundamental domain 
made of $1/12$th of the original hexagon (see Fig.\tsscppnilp), and to count them using 
Non-Intersecting Lattice Paths (NILPs). The latter are simply obtained by following successions 
of two of the three types of tiles used in the tiling of the fundamental domain. Deforming
slightly the geometry, we arrive at the problem of counting $n-1$ NILPs on the square lattice,
each path taking {\it vertical} steps of $(0,1)$ or {\it diagonal} steps of $(1,1)$ only, the $i$-th path
starting at point $(i,-i)$ and ending up on the line $y=0$, after $i$ steps.
This leads to the total number of such paths $N_{10}(2n)$, as given for instance by a 
Lindstr\"om-Gessel-Viennot (LGV) type formula \LGV, expressing the number of NILPs with fixed endpoints
as the determinant of the numbers of paths from the $i$-th starting point to the $j$-th endpoint 
say $(r_j,0)$, and summed over the positions of the endpoints.
In Ref.\DFb, a refinement of these numbers was introduced via the generating polynomial
$N_{10}(2n\vert \tau)$ of TSSCPPs counted with a weight $\tau$ per vertical step, and given 
by the following LGV type formula:
\eqn\neightau{ N_{10}(2n\vert \tau)=\sum_{1\leq r_1<r_2<\cdots <r_{n-1}\atop
r_i\leq 2i }\det_{1\leq i,j\leq n-1}
\left( {i\choose r_j-i} \tau^{2i-r_j} \right)}
\example For $n=3$, there are $7$ TSSCPP configurations:
$$\figbox{13.cm}{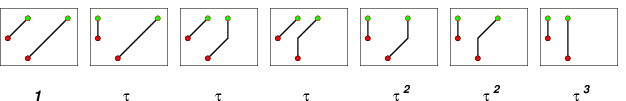}$$
resulting in the polynomial $N_{10}(6\vert \tau)=1+3\tau+2\tau^2+\tau^3$.

\subsec{Integral formulae}
Let us derive a simple multiple contour integral formula for the generating polynomial
$N_{10}(t_1,t_2,\ldots t_{n-1})$ of TSSCPPs in their NILP formulation with a weight $t_i$
per vertical step in the $i$-th slice delimited by the lines $y=1-i$ and $y=-i$. We use
again the LGV formula, including the weighting of paths by the $t_i$'s.
The formula reads 
\eqn\refinfomrutss{N_{10}(t_1,t_2,\ldots t_{n-1})=\sum_{1\leq r_1<r_2<\cdots <r_{n-1}\atop
r_i\leq 2i } \det_{1\leq i,j\leq n-1} \, {\cal P}_{i,r_j} }
where ${\cal P}_{i,r}$ is the weighted sum over all lattice paths
from $(i,-i)$ to $(r,0)$. Such a path has $r-i$ diagonal steps and $2i-r$ vertical ones,
to be taken within the $i$ first slices, hence
\eqn\valcalp{ {\cal P}_{i,r}=\sum_{1\leq i_1<i_2<\cdots <i_{2i-r}} 
\prod_{\ell=1}^{2i-r} t_{i_\ell} = \prod_{k=1}^{i} (1+t_k u)\Big\vert_{u^{2i-r}} }
where the subscript stands for the coefficient of the corresponding power of $u$
in the polynomial. Introducing an extra trivial path starting and ending at the origin,
such that ${\cal P}_{i,0}=\delta_{i,0}$
and 
substituting \valcalp\ into \refinfomrutss, we get the following coefficient identity:
\eqn\finlut{ N_{10}(t_1,t_2,\ldots t_{n-1})=
\prod_{1\leq i<j\leq n} {(u_j-u_i)(1+t_i u_j)\over 1-u_iu_j} \prod_{i=1}^{n} {1\over 1-u_i}
\Bigg\vert_{\prod_{i=1}^{n} u_i^{2i-2}}}
where again the subscript stands for the coefficient of the corresponding monomial
in the power series expansion of the rational fraction around $0$. To prove \finlut, 
one uses the multilinearity of the determinant to rewrite \refinfomrutss--\valcalp\ as
\eqn\rewtss{
N_{10}(t_1,t_2,\ldots t_{n-1})=
\prod_{i=1}^{n}\oint {\d u_i\over u_i^{2i-1}}\prod_{k=1}^{i-1} (1+t_k u_i)
\sum_{0\leq r_0< r_1<r_2<\cdots <r_{n-1}} \det_{1\leq i,j\leq n}\Big( u_i^{r_{j-1}}\Big) }
where we have replaced the condition $r_0=0$ by a summation over $r_0\geq 0$, that does not
alter the result of the integration.
Finally, we evaluate the sum of determinants to be
\eqn\sumofdet{\sum_{0\leq r_0< r_1<r_2<\cdots <r_{n-1}} \det_{1\leq i,j\leq n}\Big( u_i^{r_{j-1}}\Big)=
\prod_{1\leq i<j\leq n} {u_j-u_i\over 1-u_iu_j} \prod_{i=1}^{n} {1\over 1-u_i} }
This is nothing but the sum of all Schur functions of the $n$ arguments
$u_1,\ldots,u_{n}$, multiplied by the Vandermonde determinant. The result
for this sum is standard (see for instance eq 4.17 of \Bre). Eq. \finlut\ then follows.

\fig{NILPs for Modified TSSCPPs. We have added an extra top slice of steps to the original NILPs
of Fig\tsscppnilp, so that the new NILPs end up on the line $y=1$. 
The arrival points, of the form $(1+r_i,1)$ are further
constrained by imposing that all successive differences $r_{i+1}-r_i$ be odd
integers. With this, there is exactly one way of completing a TSSCPP into a modified one,
thus the two sets of objects are in bijection.}{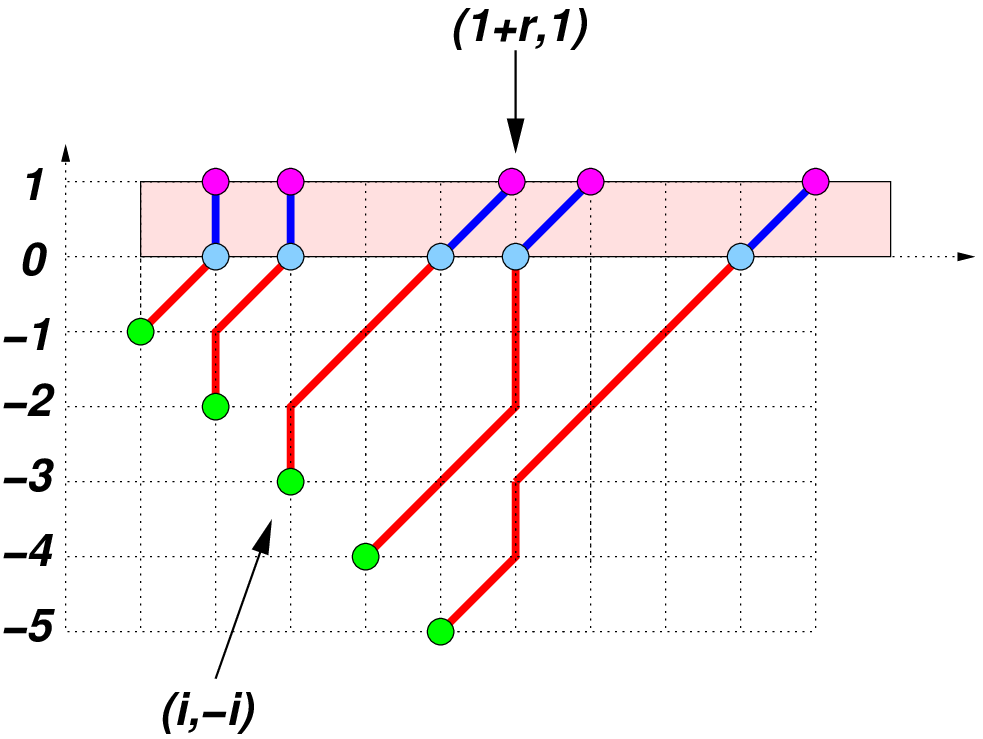}{8.cm}
\figlabel\moditsscpp

For later purposes, we are also interested in modified TSSCPPs defined as follows. 
We simply add up one extra step to all previous TSSCPPs in a top slice between $y=0$ and $y=1$,
with a weight $t_0$ per vertical step in this extra slice,
and further constrained as follows.
We demand that the consecutive new endpoints, of the form $(r_i+1,1)$, differ only by some
odd integers, namely $r_{i+1}-r_i$ odd for all $i$, and $r_1=1$. The reason for this is that
there is a bijection between these and the TSSCPPs above. Indeed, the constraint on the
new endpoints ensures that the restrictions of these modified TSSCPPs to the region
below the x axis be in bijection with regular TSSCPPs. 
We denote by $N_{10}'(t_0,t_1,\ldots,t_{n-1})$ the corresponding generating polynomial.
\example The seven TSSCPPs at $n=3$ are augmented as follows:
$$\figbox{13.cm}{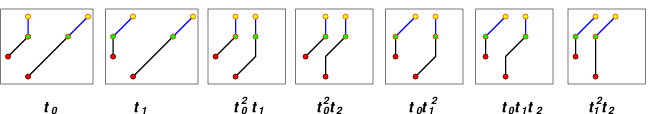}$$
and the corresponding generating polynomial reads 
$N_{10}'(t_0,t_1,t_2)=(1+t_0t_1)(t_0+t_1)+(t_0^2+t_0t_1+t_1^2)t_2$.

We have:
\eqn\moditss{ N_{10}'(t_0,t_1,\ldots,t_{n-1})=
\prod_{1\leq i<j\leq n} {(u_j-u_i)(1+t_i u_j)\over 1-u_iu_j} 
\prod_{i=1}^{n} {1+t_0 u_i\over 1-u_i^2}
\Bigg\vert_{\prod_{i=1}^{n} u_i^{2i-2}}}
As expected from the above bijection, we recover \finlut\ at $t_0=1$.
Eq.~\moditss\ is proved in much the same way as \finlut. Denoting by ${\cal Q}_{i,r}$
the weighted sum over paths from $(i,-i)$ to $(r+1,1)$, we now have
\eqn\valcalq{ {\cal Q}_{i,r}=\prod_{k=0}^i (1+t_k u)\Big\vert_{u^{2i-r}} }
and, as in the previous case, adding up a trivial path from the origin to $(1,1)$ (hence an extra
arrival point $r_0=0$), we have
\eqn\finlutmod{ N_{10}'(t_0,t_1,\ldots,t_{n-1})=
\prod_{i=1}^{n}\oint{\d u_i\over 2\pi i u_i^{2i-1}}
\prod_{k=0}^i (1+t_k u_i)\sum_{0\leq r_0<r_1<\cdots <r_{n-1}\atop r_{i+1}-r_i\ {\rm odd}} 
\det_{1\leq i,j\leq n}\Big( u_i^{r_{j-1}}\Big) }
Note that as before we have relaxed the condition $r_0=0$ into $r_0\geq 0$, with $r_0$ even
(we have introduced $r_{-1}\equiv -1$, and the condition $r_{i+1}-r_i$ odd applies for 
$i=-1,0,\ldots,n-2$).
The last sum is easily evaluated from a standard result for the sum over all Schur
functions corresponding to {\it even}\/ partitions (see eq. 4.29 of \Bre), namely that
\eqn\evenpati{ \sum_{0\leq r_0<r_1<\cdots <r_{n-1}\atop r_{i+1}-r_i\ {\rm odd}} 
\det_{1\leq i,j\leq n}\Big( u_i^{r_{j-1}}\Big)=
\prod_{1\leq i<j\leq n} {u_j-u_i\over 1-u_iu_j} 
\prod_{i=1}^{n} {1\over 1-u_i^2}}
and eq.\moditss\ follows.

The modified TSSCPP generating polynomial $N_{10}'(t_0,t_1,\ldots,t_{n-1})$
will be specialized for later purpose to $t_0\equiv t$ and $t_1=t_2=\cdots =t_{n-1}=\tau$,
and we denote by $N_{10}'(2n\vert t,\tau)$ the corresponding polynomial. For instance
in the case $n=3$, 
$N_{10}'(6\vert t,\tau)=N_{10}'(t,\tau,\tau)=t+\tau+2t^2\tau+2t\tau^2+\tau^3$.
From the remark above, we have $N_{10}'(2n\vert 1,\tau)=N_{10}(2n\vert \tau)$, as
the modified TSSCPPs are in bijection with TSSCPPs, and $t=1$ corresponds to putting no
extra weight for the modification. Hence for $\tau=1$, we have
$N_{10}'(2n\vert 1,1)=N_{10}(2n\vert 1)=A_n$. This yields for instance 
$N_{10}'(6\vert 1,1)=(t+\tau+2t^2\tau+2t\tau^2+\tau^3)_{t=\tau=1}=7=A_3$.

Interestingly, if we take $t_0=0$, the only contributing modified TSSCPPs are those
with only diagonal steps in their last row, which enforces the rule that consecutive
arrival points on the underlying TSSCPPs (with top slice removed) must have odd integer
differences. Therefore
\eqn\restrictnten{ N_{10}'(t_0=0,t_1,\ldots,t_{n-1})= N_{10}'(t_1,\ldots,t_{n-1})}
Once expressed at the above specialization, this gives
$N_{10}'(2n\vert 0,\tau)=N_{10}'(2n-2\vert \tau,\tau)$. In particular, if we take $\tau=1$,
we get $N_{10}'(2n\vert 0,1)=N_{10}'(2n-2\vert 1,1)=N_{10}(2n-2\vert 1)=A_{n-1}$. For instance,
$N_{10}'(6\vert 0,1)=(t+\tau+2t^2\tau+2t\tau^2+\tau^3)\vert_{t=0,\tau=1}=2=A_2$.
Another simple specialization is when $t_0\to \infty$, in which case we retain only
the configurations with only vertical steps in the extra slice, hence again such that
the endpoints $(r_i,0)$ of the corresponding TSSCPPs themselves satisfy the parity condition
that all $r_{i+1}-r_i$ are odd integers:
\eqn\specinf{\lim_{t_0\to \infty} t_0^{-(n-1)} N_{10}'(t_0,t_1,\ldots,t_{n-1})
=N_{10}'(t_1,\ldots,t_{n-1})}
When all $t_i=1$ for $i=1,2\ldots,n-1$, this
is nothing but $N_{10}'(2n-2\vert 1,1)=A_{n-1}$. In the case $n=3$, this reduces to
$(t+\tau+2t^2\tau+2t\tau^2+\tau^3)\vert_{t^2}\vert_{\tau =1}=2=A_2$.

The two specializations at $t_0=0$ and $t_0=1$ will correspond precisely to the 
cases (ii) and (iii) described in the introduction. For $t$ arbitrary and $\tau=1$,
we also get a refinement of the TSSCPP numbers in the form of a polynomial
$N_{10}'(2n\vert t,1)$. The coefficients of this polynomial seem to be the
refined alternating sign matrix numbers $A_{n,k}$ that count alternating
sign matrices of size $n\times n$, with a $1$ in position $k$ on their top row.
So we have the conjecture that:
\eqn\reftsscpp{ N_{10}'(2n\vert t,1)=\sum_{k=1}^n A_{n,k}t^{k-1} }
Note that an analogous statement \ROB\ exists
for refined TSSCPPs of size $2n$ according to the
steps in their top slice, namely
\eqn\protss{ N_{10}(t,1,1,\ldots,1)=\sum_{k=1}^n A_{n,k}t^{k-1} }

\newsec{The minimal polynomial solution to the $q$KZ equation}

\subsec{Level $1$ $q$KZ equation}
The $q$KZ equation was introduced in the context of affine quantum
groups \FR\ and in connection to two-dimensional integrable lattice
models \Smi.
Our basic observation \DFZJc\ is that it allows for generalizing
the groundstate eigenvector condition \evectP\ to generic values of $q$.
Here we concentrate on the level 1 
$U_q(\widehat{\goth sl_2})$ $q$KZ equation, and
refer to \DFZJc\ for higher rank generalizations. The latter is based on the 
Temperley-Lieb algebra $TL_{2n}(\tau)$ with generators $e_i$, $i=1,\ldots,2n$
subject to the relations: $e_i^2=\tau e_i$ and $e_i e_{i\pm 1 } e_i=e_i$ for all $i$,
with the convention that $e_{2n+1}\equiv e_1$.
The standard representation of $TL_{2n}(\tau)$ is on link patterns $\pi\in \LP_{2n}$.
The generator $e_i$ simply acts by connecting the two points $i,i+1$, and gluing the 
two other ends of arches formerly attached to them. If $i$ and $i+1$ were already 
connected, $e_i$ acts as the identity times the parameter $\tau$, which accounts
for a weight for the loop thus created. This weight per loop is parametrized as
\eqn\weightperloop{ \tau=-q-q^{-1} }
where we might interpret $-q^\epsilon$ as a weight per oriented loop, $\epsilon=\pm 1$
according to the orientation, and the weight $\tau$ is obtained by summing over the 
two possible orientations of each loop.

Using the Temperley-Lieb algebra generators, it is a simple exercise to construct
a solution of the unitarity and Yang-Baxter equations, in the form of 
the quantum $R$-matrix:
\eqn\rmat{\Rc_{i,i+1}(z,w)={q\,z-q^{-1}w\over q\,w-q^{-1}z} \, 
{\rm Id}+{z-w\over q\,w-q^{-1}z} \, e_i }

The $q$KZ equation reduces to the following relations:
\eqn\qkzred{\eqalign{ \Rc_{i,i+1}(z_{i+1},z_{i}) \, \Psi(z_1,\ldots,z_{2n})
&= \tau_i \Psi(z_1,\ldots,z_{2n}), \quad i=1,2,\ldots,2n-1\cr
\sigma\Psi(z_2,\ldots,z_{2n},s z_1)&= c_n \Psi(z_1,\ldots,z_{2n})\cr}}
where the operator $\tau_i$ acts as the elementary transposition of
spectral parameters $z_i\leftrightarrow z_{i+1}$ and the operator $\sigma$ is a
shift operator $\sigma e_i \sigma^{-1}=e_{i+1}$ for all $i$, while the
constants $s$ and $c_n$ are fixed to be respectively $s=q^6$ and $c_n=q^{3(n-1)}$.

Expressing \qkzred\ in the link pattern basis, we get a system of equations
determining the components $\Psi_\pi$ of the vector $\Psi_n$, namely:
\eqn\compopsi{ \eqalign{
{q^{-1}z_{i+1}-q z_i\over z_{i+1}-z_i} (\tau_i -1)\Psi_\pi(z_1,\ldots,z_{2n}) 
&=\sum_{\pi'\neq \pi\atop e_i \pi'=\pi} \Psi_{\pi'}(z_1,\ldots,z_{2n}) \cr
\Psi_{\sigma(\pi)}(z_2,\ldots,z_{2n},q^6 z_1)
&=q^{3(n-1)} \Psi_\pi(z_1,\ldots,z_{2n})\cr}}

In \refs{\Pas,\KP}, it was shown that the space, called $V$ in what
follows, which is spanned by the $\Psi_\pi$ could be
simply characterized by the following vanishing condition (also called
wheel condition), that any polynomial in that space vanishes
whenever any triple of cyclically ordered spectral parameters take values 
$(z, q^2 z,q^4 z)$. We will make use of this remark below.

Let us quote the explicit form of the base component $\Psi_{\pi_0}$ corresponding
to the link pattern $\pi_0$ that connects $i$ and $N+1-i$:
\eqn\basecase{
\Psi_{\pi_0}
=\prod_{1\le i<j\le n} (q\,z_i-q^{-1}z_j)
\prod_{n+1\le i<j\le N} (q\,z_i-q^{-1}z_j)
}
as well as a useful property that allows to generate $\Psi_n$ recursively:

{\sl The components of $\Psi_n$, polynomial solution of 
level $1$ $q$KZ equation
of degree $n(n-1)$, normalized by Eq.~\basecase,
satisfy the following recurrence relations:
\eqn\recurcompeq{
\Psi_\pi|_{z_{i+1}=q^2 z_i}=\cases{
0& if $\pi(i)\ne i+1$.\cr
\prod_{j=1}^{i-1}(z_i-q^2 z_j) \prod_{j=i+2}^N (q^2 z_i-q^{-2}z_j)\times
&if $\pi(i)=i+1$.
\cr\times \Psi_{\pi'}(z_1,\ldots,z_{i-1},
z_{i+2},\ldots,z_N)
\cr}
}
where in the second case, $\pi'$ is obtained from $\pi$ by deleting the arch $(i,i+1)$.}

\subsec{Integral formulae for the span of solutions}
We now turn to the derivation of multiple residue integral formulas for vectors
in the space spanned by the components $\Psi_\pi$ of the minimal polynomial solution to 
the $q$KZ equation.

Let $z_1,\ldots,z_N$ be $N=2n$ complex indeterminates (spectral parameters).
We consider the following multiple contour integrals:
\eqnn\defpsi
$$\eqalignno{
\Psi_{a_1,\ldots,a_n}&=
\prod_{1\le i<j\le N}(q\,z_i-q^{-1}z_j)\times\cr
&\times \oint\cdots \oint
\prod_{\ell=1}^n {\d w_\ell\over 2\pi i}
{
\prod_{1\le \ell<m\le n} (w_m-w_\ell)(q\,w_\ell-q^{-1}w_m)
\over
\prod_{\ell=1}^n \prod_{1\le i\le a_\ell} (w_\ell-z_i) 
\prod_{a_\ell<i\le N} (q\,w_\ell-q^{-1}z_i)
}
&\defpsi}$$

where $(a_\ell)_{\ell=1,\ldots,n}$ is any {\it non-decreasing}\/ sequence of integers in 
$\{1,\ldots,N-1\}$.
The contours catch the poles at $w_i=z_j$ but not those at $w_i=q^{-2}z_j$.
These integrals are closely related to formulae for solutions of level
1 $q$KZ equation in the {\it spin}\/ basis, as given in e.g.~\JM. In
appendix B, a detailed discussion of the connection between the two
types of integrals is given.

We want to show the following:
{\sl $\Psi_{a_1,\ldots,a_n}$ is a homogeneous polynomial in the variables $z_1,\ldots,z_N$
of degree $n(n-1)$.
Furthermore it satisfies the wheel condition: for all ordered
triplets $i,j,k$,
\eqn\vancond{
\Psi_{a_1,\ldots,a_n}(\ldots,z_i=z,\ldots,z_j=q^2 z,
\ldots,z_k=q^4 z,\ldots)=0\qquad 1\le i<j<k\le N}}

To prove this, we first write explicitly the residue formula for Eq.~\defpsi:
\eqnn\respsi
$$\eqalignno{
&\Psi_{a_1,\ldots,a_n}=\prod_{1\le i<j\le N}(q\,z_i-q^{-1}z_j)\times\cr
&\qquad\qquad\times
\!\!\sum_{\{k_1,\ldots,k_n\}\atop k_\ell\ne k_m,1\le k_\ell\le a_\ell}
{\prod_{1\le \ell<m\le n} (z_{k_m}-z_{k_\ell})(q\,z_{k_\ell}-q^{-1}z_{k_m})
\over
\prod_{\ell=1}^n \prod_{1\le i\le a_\ell,i\ne k_\ell} (z_{k_\ell}-z_i) 
\prod_{a_\ell<i\le N} (q\,z_{k_\ell}-q^{-1}z_i)
}
\cr
&=
\!\!\sum_{K=\{k_1,\ldots,k_n\} \atop k_\ell\ne k_m, 1\le k_\ell\le a_\ell} (-1)^{s(k_\ell)}
{\displaystyle
\prod_{1\le \ell<m\le n} (q\,z_{k_\ell}-q^{-1}z_{k_m})
\prod_{1\le i<j\le N\atop i\not\in K\ {\rm or}\ i=k_\ell, j\le a_\ell}(q\,z_i-q^{-1}z_j)
\over
\displaystyle
\prod_{\ell=1}^n \prod_{1\le i\le a_\ell\atop i\not\in K\ {\rm or}\ i>k_\ell} (z_{k_\ell}-z_i) 
}&\respsi
}$$
where $(-1)^{s(k_\ell)}$ is the sign of the permutation that places the $k_\ell$ in
increasing order.

Let us now compute the residue at $z_i\to z_j$. Note that at least one
of the two integers $i$, $j$ must belong to $K$ for the residue 
of the summand to be non-zero.
Two cases arise: a) terms where both $i$ and $j$ are in $K$,
say $j=k_\ell$ and $i=k_m$ with $k_\ell<k_m\le a_\ell$ (and as always
$k_m\le a_m$).
Then one can switch $k_\ell$ and $k_m$: now $k_\ell=i$, $k_m=j$,
$k_m<k_\ell\le a_m$ so this new term also has a pole at $z_i\to z_j$, the
residue being the same but with the sign changed (due to $s(k_\ell)$). 
So the two terms cancel. b) terms where
only say $j=k_\ell$, and $i\not\in K$, with $i\le a_\ell$. Consider the term where
$k_\ell=j$ is replaced with $i$: again it is tedious but easy to check that
it has the same residue with the opposite sign.

$\Psi_{a_1,\ldots,a_n}$, being a rational fraction without poles, is 
a polynomial. The homogeneity and degree follow immediately. Consider now
$1\le i<j<k\le N$, and $z_i=z$, $z_j=q^2 z$, $z_k=q^4 z$.
For each term in the sum of Eq.~\respsi: for the second type of factors
in the numerator to be non-zero, necessarily $i\in K$ and $j\in K$.
Furthermore, for the first type of factors to be non-zero,
$i=k_m$ and $j=k_\ell$ with $\ell<m$. But then $j=k_\ell\le a_\ell\le a_m$ (this is
where we use the fact that $(a_\ell)$ is non-decreasing), so that the second
factor with indices $i,j$ vanishes. All terms vanishing, the
sum is zero. 

\example For $\Psi_{a_1,\ldots,a_n}$ to be non-zero, according to the residue formula \respsi,
one must have $a_\ell\ge \ell$ (in fact, $\ell\le a_\ell<\ell+n$).
Let us compute the first non-trivial component, that is $\Psi_{1,2,\ldots,n}$. 
There is a single residue,
at $w_i=z_i$, $i=1,\ldots,n$, and we obtain the ``base'' component
$\Psi_{\pi_0}$ given by Eq.~\basecase.

The precise statement found in \Pas\ is that 
the space $V$ of homogeneous polynomials of degree $n(n-1)$
in the variables $z_1,\ldots,z_N$
that satisfy the condition \vancond\ is of dimension 
$c_n=(2n)!/n!/(n+1)!$, and that the components $\Psi_\pi$
of the level $1$ solution
of $q$KZ equation in the link pattern basis form a basis of this space (the components
in the spin basis also span this space, but they are not
linearly independent since there are ${2n\choose n}$ of them).
Furthermore, we have the following proposition:
{\sl a polynomial $P\in V$ is entirely determined by its values
at the following $c_n$ points, indexed by link patterns $\pi$:
\eqn\evalz{
z_i=q^{-\epsilon_i(\pi)}\qquad 
\epsilon_i(\pi)={\rm sign}(\pi(i)-i)=
\cases{
+1&if $\pi$ has an opening at $i$\cr
-1&if $\pi$ has a closing at $i$\cr
}}}

Proof. Write $P=\sum_\pi a_\pi \Psi_\pi$, and note that
$\Psi_{\pi'}(q^{-\epsilon_i(\pi)})\ne 0$ iff $\pi=\pi'$. This is easily
proved by induction using Eq.~\recurcompeq. Indeed pick any ``little arch'' of $\pi$,
that is $i$ such that $\pi(i)=i+1$. Either 
a) $\pi'(i)\ne i+1$, in which case $\Psi_{\pi'}(q^{-\epsilon_i(\pi)})=0$ (first case
of Eq.~\recurcompeq);
or b) $\pi(i)=i+1$, in which
case we may apply the second case of Eq.~\recurcompeq\ 
and use the induction hypothesis to conclude.
The induction even allows to compute $C_\pi:=\Psi_\pi(q^{-\epsilon_i(\pi)})$,
though we shall not need these explicit expressions:
\eqn\cons{
C_\pi=C\,\tau^{|\pi|}\qquad C=(q-q^{-1})^{n(n-1)}\qquad
|\pi|=n^2+\sum_{i=1}^N i\, \epsilon_i(\pi)
}
Going back to $P$, we find that $a_\pi=P(q^{-\epsilon_i(\pi)})/C_\pi$, and in
particular that $P$ is entirely determined by these values. 

Each $\Psi_{a_1,\ldots,a_n}$ is thus a linear combination of the $\Psi_\pi$,
and the coefficients are given by a simple evaluation. In the course of this
proof we have found that the evaluation of the $\Psi_\pi$
themselves is most easily obtained by using recurrence relations; it is
therefore natural to try to do the same for the $\Psi_{a_1,\ldots,a_n}$:

{\sl Assume that $(a_i)$ is an {\it increasing}\/ sequence. Then
\eqn\receqa{
\Psi_{a_1,\ldots,a_n}|_{z_{i+1}=q^2 z_i}=\cases{
0&\hskip-1cm if $i\not\in \{ a_\ell \}$.\cr
\prod_{j=1}^{i-1}(z_i-q^2 z_j) \prod_{j=i+2}^N (q^2 z_i-q^{-2}z_j)\times
&\hskip-1cm if $i=a_\ell$.
\cr \times\Psi_{a_1,\ldots,a_{\ell-1},a_{\ell+1}-2,\ldots,a_n-2}(z_1,\ldots,z_{i-1},
z_{i+2},\ldots,z_N)
\cr}
}}

Note that the sequence $(a_1,\ldots,a_{\ell-1},a_{\ell+1}-2,\ldots,a_n-2)$ is
not necessarily increasing (only non-decreasing).

Proof. Assume $z_{i+1}=q^2 z_i$. According to Eq.~\respsi, 
for the second type of factor in the numerator to be non-zero, $i\in K$,
so that $i=k_\ell$ for some $\ell$, and $i+1>a_\ell$. Since $k_\ell\le a_\ell$,
$i=k_\ell=a_\ell$. This proves the first case of Eq.~\receqa. 
For the second case, we see that the $\ell$ above being unique,
only the
residue at $w_\ell=z_i$ contributes, so that we can go back to Eq.~\defpsi\ and
perform the integration over $w_\ell$. We find:
\eqnn\receq
$$\eqalignno{
\Psi_{a_1,\ldots,a_n}&=
\prod_{1\le j<k\le N, (j,k)\ne (i,i+1)}(q\,z_j-q^{-1}z_k)
\oint\cdots\oint \prod_{1\le m\le n, m\ne\ell} {\d w_m\over 2\pi i}\cr
&
{
\prod_{1\le m<m'\le n} (w_{m'}-w_m)(q\,w_m-q^{-1}w_{m'})
\over
\prod_{1\le m\le n,m\ne\ell}^n \prod_{1\le j\le a_m} (w_m-z_j) 
\prod_{a_m<j\le N} (q\,w_m-q^{-1}z_j)}\cr
&{\prod_{1\le m<\ell}(z_i-w_m)(q\,w_m-q^{-1}z_i)\prod_{\ell<m\le n}(w_m-z_i)(q\,z_i-q^{-1}w_m)
\over \prod_{1\le j<i} (z_i-z_j)\prod_{i+1<j\le N} (q\, z_i-q^{-1}z_j)}
&\receq}
$$
Using $z_{i+1}=q^2 z_i$ results in multiple cancellations, and after pulling out 
the factors appearing in
$\Psi_{a_1,\ldots,a_{\ell-1},a_{\ell+1}-2,\ldots,a_n-2}$ with appropriately reindexed variables, we get
Eq.~\receqa.

This proposition may be generalized to arbitrary sequences, see
appendix A.

\subsec{Partial sums in the link pattern basis}
We now consider a special class of such integrals,
corresponding to the following increasing sequences $(a_\ell)$:
\eqn\defA{
{\cal A}_n=\{ (a_\ell)_{1\le\ell\le n} : a_\ell\in \{2\ell-2,2\ell-1\}\}
}
This defines $2^{n-1}$ different sequences ($a_1=1$; one could consider $a_1=0$ but it would
correspond to $\Psi_{0,a_2,\ldots,a_n}=0$). One interesting property of these
is that the recurrence relation \receqa\ expresses sequences from ${\cal A}_n$ in terms of sequences from ${\cal A}_{n-1}$. 
We also consider a partition
of the set of link patterns into subsets indexed by the same sequences $(a_\ell)\in {\cal A}_n$:
\eqn\defL{
{\cal L}(a_1,\ldots,a_n)=\{ \pi\in\LP_{2n}: \pi(2m-1)>2m-1\ \hbox{iff}\ 2m-1\in \{ a_\ell \}, m=1,\ldots,n \}
}
i.e.\ the set of link patterns whose arch openings on odd sites are exactly the odd elements
of the corresponding sequence.

We can now state the following identities:
\eqn\partsumeq{
\Psi_{a_1,\ldots,a_n}=\sum_{\pi\in {\cal L}(a_1,\ldots,a_n)} \Psi_\pi
\qquad\forall (a_\ell)\in {\cal A}_n
}

Proof. Both sides of Eq.~\partsumeq\ belong to $V$.
It thus suffices to show that they are equal at the $c_n$ values of
Eq.~\evalz. To perform this evaluation for the l.h.s.
it is enough to use Eq.~\receqa, i.e.\ use recurrence relations
at $z_{i+1}=q^2 z_i$: indeed, to evaluate at
$z_i=q^{-\epsilon_i(\pi)}$, one can pick a little arch $(i,i+1)$ of $\pi$ and
note that $z_{i+1}=q=q^2\times q^{-1}=q^2 z_i$, so that one can apply
the recurrence relation, which reduces to the same evaluation (with
link pattern obtained from $\pi$ by removal of the little arch $(i,i+1)$)
in size $n-1$.
Thus, one must show the same recurrence relations for the r.h.s. 
We shall use Eq.~\recurcompeq. Set $z_{i+1}=q^2 z_i$, and assume first $i\not\in \{a_\ell\}$.
If $i$ is odd that means all $\pi\in {\cal L}(a_\ell)$ do not have an opening at $i$.
If $i$ is even then $i+1\in \{a_\ell\}$ (cf Eq.~\defA) and thus all $\pi\in {\cal L}(a_\ell)$ have
an opening at $i+1$. In both cases we conclude that they have no little arch between $i$ and
$i+1$ and thus are zero, so that the r.h.s.\ of Eq.~\partsumeq\ is zero.
Assume now $i=a_\ell$. Among the $\pi\in {\cal L}(a_1,\ldots,a_n)$, there are two categories:
those with $\pi(i)\ne i+1$, which do not contribute to the sum at $z_{i+1}=q^2 z_i$; and
those with $\pi(i)=i+1$. It is easy to check that the latter are in bijection
with the $\pi'\in {\cal L}(a_1,\ldots,a_{\ell-1},a_{\ell+1}-2,\ldots,a_n-2)$, the bijection consisting
as usual in the deletion of the little arch $(i,i+1)$. Since the prefactors in Eqs.~\recurcompeq\ and
\receqa\ are the same, we conclude that the recurrence relations satisfied by both sides of 
Eq.~\receqa\ are identical. The initial condition at $n=0$ is also identical: $\Psi_\emptyset=1$.

\example At $n=3$, ${\cal A}_3=\{(1,3,5),(1,3,4),(1,2,5), (1,2,4)\}$ and the partition of link patterns is
$$\eqalignno{
{\cal L}(1,3,5)&=\Big\{ \raisedfigbox{4cm}{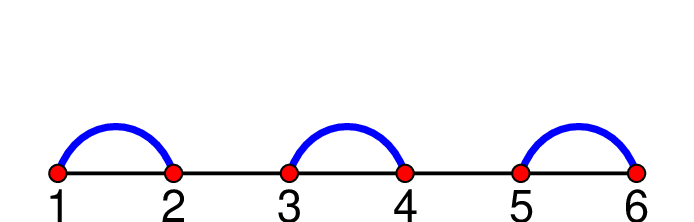}{-0.2cm} \Big\}\cr
{\cal L}(1,3,4)&=\Big\{ \raisedfigbox{4cm}{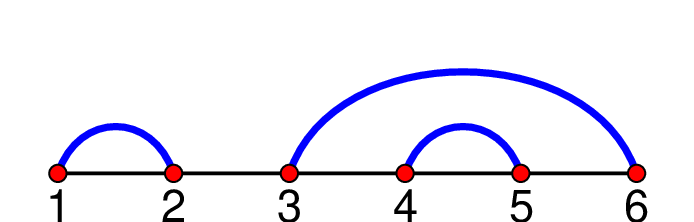}{-0.2cm}, \raisedfigbox{4cm}{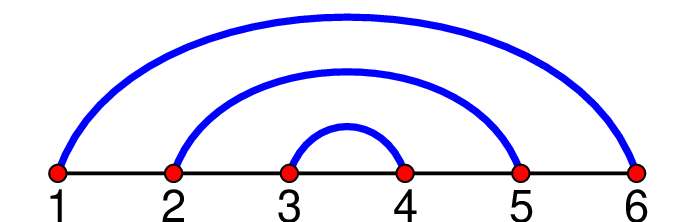}{-0.2cm} \Big\}\cr
{\cal L}(1,2,5)&=\Big\{ \raisedfigbox{4cm}{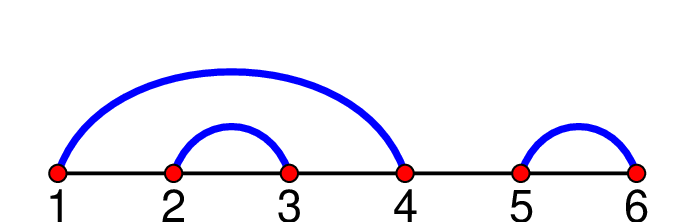}{-0.2cm} \Big\}\cr
{\cal L}(1,2,4)&=\Big\{ \raisedfigbox{4cm}{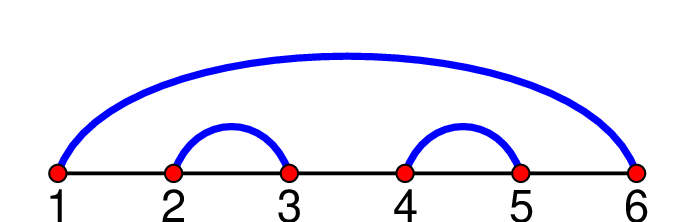}{-0.2cm} \Big\}\cr
}$$

Note that ${\cal L}(1,3,\ldots,2n-1)$ is always the singlet $\pi(2\ell-1)=2\ell$, $\ell=1,\ldots,n$ and
that ${\cal L}(1,2,\ldots,2n-2)$ is always the singlet $\pi(1)=N$, $\pi(2\ell)=2\ell+1$, $\ell=1,\ldots,n-1$.
In general, ${\cal L}(a_1,\ldots,a_n)$ contains the link pattern whose openings are exactly the $a_\ell$, but may
have more elements.

\subsec{Homogeneous limit}
Finally we consider the homogeneous limit $z_1=\cdots=z_{2n}=1$; 
We rewrite Eq.~\defpsi\ with this specialization and normalize it with $\Psi_{\pi_0}=(q-q^{-1})^{n(n-1)}$:
\eqn\coroleqa{
\Psi_{a_1,\ldots,a_n}/\Psi_{\pi_0}=
(q-q^{-1})^{n^2} \oint\cdots\oint \prod_{\ell=1}^n {\d w_\ell\over 2\pi i}
{\prod_{1\le \ell<m\le n} (w_m-w_\ell)(q\, w_\ell-q^{-1}w_m)
\over \prod_{\ell=1}^n (w_\ell-1)^{a_\ell} (q\,w_\ell-q^{-1})^{2n-a_\ell}
}
}
Changing variables: $w_\ell={1-q^{-1}u_\ell\over 1-q\,u_\ell}$ results in
\eqn\coroleq{
\Psi_{a_1,\ldots,a_n}/\Psi_{\pi_0} = 
\oint\cdots\oint \prod_{\ell=1}^n {\d u_\ell\over 2\pi i
\, u_\ell^{a_\ell}}
\prod_{1\le \ell<m\le n} (u_m-u_\ell)(1+u_\ell u_m+\tau u_m)
}
where the contours surround $0$, and $\tau=-q-q^{-1}$. 

As a corollary, the sum of all components is
\eqn\coroleqsum{
\sum_{\pi\in \LP_{2n}} \Psi_\pi/\Psi_{\pi_0} = 
\oint\cdots\oint \prod_{\ell=1}^n {(1+u_\ell)\d u_\ell\over 2\pi i
\, u_\ell^{2\ell-1}}
\prod_{1\le \ell<m\le n} (u_m-u_\ell)(1+u_\ell u_m+\tau u_m)
}
This is obtained by summing Eq.~\coroleq\ over all sequences $(a_\ell)\in {\cal A}_n$
and applying Eq.~\partsumeq.

\example At $n=3$, $\Psi_{1,3,5}=\tau^3+\tau$, $\Psi_{1,3,4}=\tau^2+1$, 
$\Psi_{1,2,5}=\tau^2$, $\Psi_{1,2,4}=2\tau$ and the full sum is $\sum_\pi \Psi_\pi=\tau^3+2\tau^2+3\tau+1$.

\newsec{From integral formulae back to TSSCPPs}
\subsec{A refined $q$KZ--TSSCPP conjecture}
Let us consider once again the generating series $N'_{10}$ of modified TSSCPPs specialized
to $t_0=t$ and $t_1=t_2=\cdots=t_{n-1}=\tau$: according to Eq.~\moditss, it is given by
\eqn\inta{
N'_{10}(2n | t,\tau) =\oint\cdots\oint\prod_{\ell=1}^n 
{\d u_\ell(1+t u_\ell)(1+\tau u_\ell)^{\ell-1}\over 2 \pi i u_\ell^{2\ell-1}}
{\prod_{1\le \ell<m\le n} (u_m-u_\ell)
\over
\prod_{1\le \ell\le m\le n} (1-u_\ell u_m)}}
On the other hand, in view of the partial sums of 
Eq.~\partsumeq, it is natural to define the following generating series:
\eqn\intb{
\hat N'_{10}(2n|t,\tau):=
\sum_{(a_\ell)\in A} t^{\sum_\ell (2\ell-1-a_\ell)} \sum_{\pi\in {\cal L}(a_1,\ldots,a_n)} \Psi_\pi/\Psi_{\pi_0}
}
Applying Eq.~\coroleq, we find the following formula:
\eqn\intc{
\hat N'_{10}(2n|t,\tau)= \oint\cdots\oint\prod_{\ell=1}^n{\d u_\ell(1+t u_\ell)\over 2 \pi i u_\ell^{2\ell-1}}
\prod_{1\le \ell<m\le n} (u_m-u_\ell)(1+\tau u_m+u_\ell u_m)}
Note that at $t=1$ $\hat N'_{10}(2n|1,\tau)$
is simply the sum of all components in the loop basis. The content of the TSSCPP conjecture of
\DFb\ is thus that $\hat N'_{10}(2n|1,\tau)=N'_{10}(2n|1,\tau)$.

Observe further that at $t=0$, $\hat N'_{10}(2n|0,\tau)$ is the largest component
$\Psi_{max}/\Psi_{\pi_0}$ whereas $N'_{10}(2n|0,\tau)=N'_{10}(2n-2|\tau,\tau)$. In particular at
$\tau=1$ they both equal $A_{n-1}$.

In view of these specializations and of numerical experimentation,
we have been led to the 
conjecture that $N'_{10}(2n|t,\tau)=\hat N'_{10}(2n|t,\tau)$ for all
values of the parameters. 
For example, at $n=3$, $\hat N'_{10}(6|t,\tau)=\Psi_{1,3,5}+t(\Psi_{1,3,4}+\Psi_{1,2,5})+t^2 \Psi_{1,2,4}
=\tau^3+2t\tau^2+2t^2\tau+\tau+t$, which coincides with $N'_{10}(6|t,\tau)$ given in Sect.~2.

\subsec{Attempted proof and a conjectured identity}
In order to show this conjecture, all one needs to do is to prove the equality of Eqs.~\inta\ and
\intc. Considering the free fermionic nature of TSSCPPs, it is natural to antisymmetrize these
expressions. We find the following formulae:
\eqn\damnint{
\eqalign{
&\left\{\prod_{1\le\ell\le m\le n}(1-u_\ell u_m)\ 
{\rm AS}\left(\prod_{i=1}^n u_\ell^{-2\ell+2} \prod_{1\le\ell<m\le n}
(1+u_\ell u_m+\tau u_m)\right)\right\}_{\le 0}\cr
&={\rm AS}\left(\prod_{\ell=1}^n
\left(u_\ell^{-1}(\tau +u_\ell^{-1})\right)^{\ell-1}\right)=\prod_{1\le\ell<m\le n}(u_m^{-1}-u_\ell^{-1})
(\tau+u_\ell^{-1}+u_m^{-1})\cr}}
where AS stands for the antisymmetrization with respect to permutations of
the $u$'s, namely 
$AS(f(u_1,\ldots,u_n))=\sum_{\sigma\in {\cal S}_n} (-1)^\sigma f(u_{\sigma(1)},\ldots,u_{\sigma(n)})$,
and the subscript $\le0$ means keeping
only terms with negative or zero powers in all the variables $u_\ell$. The equality in the second line
is elementary, but the antisymmetrization in the first is non-trivial. 
It is easy to check that Eq.~\damnint\ implies that the symmetrized integrands 
of Eqs.~\inta\ and \intc\ are equal, so that the integrals are equal.

Equivalently, we can rewrite Eq.~\damnint\ as the following integral identity:
\eqnn\integident
$$\eqalignno{
\oint\cdots\oint\prod_{\ell=1}^n 
{\d u_\ell\over 2\pi i}
{1-x u_\ell^2\over 2\pi i u_\ell^{2\ell}}\det_{1\leq \ell,m\leq n}
&\left( {1
\over 1-\alpha_\ell u_m}\right) \prod_{1\leq \ell<m\leq n} (1+\tau u_m+x u_\ell u_m)(1-xu_\ell u_m) \cr
&=\prod_{\ell=1}^n \alpha_\ell
\prod_{1\leq \ell<m\leq n} (\alpha_m-\alpha_\ell)(\tau+\alpha_\ell+\alpha_m)&\integident\cr}$$

There are two important things about this identity. The determinant in the integrand 
is, up to a Vandermonde determinant of the $u$'s,
the generating function for Schur polynomials of the $u$'s, themselves spanning the space 
of symmetric polynomials of these variables. Moreover, the result is clearly independent of
the parameter $x$. 
Combining these two facts, we see that if we replace 
the determinant by any symmetric polynomial of the $u$'s times the Vandermonde determinant of the 
$u$'s the integral remains independent of $x$. 
With an appropriate choice of symmetric function and by equating the $x=1$ and $x=0$ values
we recover the equality of Eqs.~\inta\ and \intc. 

\subsec{Prospects}
The great similarity of Eqs.~\inta\ and \intc\ is extremely suggestive. However these formulae are also reminiscent
of those of \Zei, which means that proving they are equal might be more difficult than it seems. 
We shall end with a few general comments.

First note that the main conjectured result of this paper, 
expounded in Sect. 4.1, relates four different kinds of 
objects: at generic parameter $\tau$, it relates 
the $\tau$-weighted counting of TSSCPPs with an additional weight $t$ 
depending on the parity of their endpoints (or equivalently,
of augmented TSSCPPs with a special weight for the last line) 
and the sum of components of the polynomial 
solution to the level $1$ $q$KZ equation (with $\tau=-q-q^{-1}$) in the link pattern basis, 
with an extra weight $t$ depending on the parity of endpoints of loops. 
When specialized to $\tau=1$, it also relates these to the $t$-refined enumeration of ASMs and
to the sum of components of the ground state of the Temperley--Lieb model of loops in which all spectral
parameters but one are taken to $1$, the latter playing the role of $t$.

None of these connections are fully understood. The introduction of spectral parameters, the main idea of \DFZJ,
has shed some light on the matter, but it is not obvious how to reintroduce a full set of parameters into the
TSSCPP conjectures. The first obvious idea that might come to mind is to use the integrability of the rhombus
tiling model, and decorate it with spectral parameters, but our attempts were unsuccessful. Note however
that for instance such a connection between the partition function for the inhomogeneous model of
rhombus tilings of an hexagon of shape $a\times b\times c$ and  the actual
components of $\Psi$ corresponding to link patterns with three sets of nested arches
was established in \PZRS. 

The techniques of the present paper may be generalized to other boundary conditions as well. For instance,
it was conjectured in \DFop\ that the sum of components of the solution to the level $1$ qKZ equation
with open (reflecting) boundaries equates in the homogeneous and generic $q$ limit the generating polynomial
for $\tau$-weighted (possibly modified) Cyclically Symmetric Transpose Complementary Plane Partitions (CSTCPP), 
namely cyclically and reflection-invariant rhombus tilings of a hexagon  
(possibly with a central triangular hole). 
More general boundaries were considered in \DFZJd, and we believe 
that integral formulae, analogous to those derived in the present paper, must exist for various sum 
rules of the components of the polynomial solution $\Psi$ to the corresponding qKZ equations, probably
allowing for a relation to $\tau$-enumeration of plane partitions with suitable symmetries.
We might also wonder how this may generalize to the higher rank and higher spin generalizations of \DFZJc\ and 
\PZspin. Note however that in the higher rank/spin cases, the generalized RS sum rules found lead to 
sequences of integer numbers for which no combinatorial interpretation exists yet, and it is an
open challenge to find out what generalization of ASM/FPL or TSSCPP they could correspond to.

Finally, one can note the great similarity of the present work with the 
recent paper \RSc. This will be discussed elsewhere \RSZJ.
\bigskip\goodbreak
\centerline{\sl Note added}
After this article was completed, Doron Zeilberger found a proof of Eq.~\damnint, which can be 
found on his web site: 
{\tt http://www.math.rutgers.edu/$\sim$zeilberg/pj.html}

\bigskip\goodbreak
\centerline{\bf Acknowledgments}
The authors acknowledge the support
of European Marie Curie Research Training Networks ``ENIGMA'' MRT-CT-2004-5652, 
``ENRAGE'' MRTN-CT-2004-005616, ESF program ``MISGAM''
and of ANR program ``GIMP'' ANR-05-BLAN-0029-01. PZJ wishes to thank the organizers of the conference
CQIS'07, where this work was completed.

\listrefs

\appendix{A}{Computing coefficients of the $\Psi_{a_1,\ldots,a_n}$ in the link pattern basis}
Eq.~\receqa, which provides recurrence relations for
 $\Psi_{a_1,\ldots,a_n}$
for $(a_\ell)$ an increasing sequence, can in fact be generalized to
arbitrary non-decreasing sequences:
\eqnn\receqplus
$$\eqalignno{
\Psi_{a_1,\ldots,\underbrace{\scriptstyle i,\ldots,i}_k,\ldots,a_n}|_{z_{i+1}=q^2 z_i}&=
\prod_{j=1}^{i-1}(z_i-q^2 z_j) \prod_{j=i+2}^N (q^2z_i-q^{-2}z_j)\times&\receqplus\cr
&\times {q^k-q^{-k}\over q-q^{-1}}\ 
\Psi_{a_1,\ldots,\underbrace{\scriptstyle i-1,\ldots,i-1}_{k-1},\ldots,a_n-2}(z_1,\ldots,z_{i-1},
z_{i+2},\ldots,z_N)
\cr}$$
The proof is along the same lines as that of Eq.~\receqa.

As a consequence, if one wants to compute the coefficients of $\Psi_\pi$ in the link pattern basis
by using evaluation of Eq.~\evalz,
the use of Eq.~\receqplus\ leads to the following:
first the non-zero coefficients correspond 
to link patterns whose little arches open in $\{a_\ell\}$; for each of
them, the computation by recurrence involves 
removing the (say, leftmost) little arch at each step; the number $k$
marked on the opening determines the coefficient 
$U_{k-1}=(q^{k}-q^{-k})/(q-q^{-1})$, and the sum of both numbers on
opening and closing minus 1 is added to the site at its left. 

It is convenient to describe graphically this process as follows. 
Shift the $a_i$ by one half-step to the right,
and put a little marker (red circle) indicating the number of 
$a_\ell$ at that location. The recurrence process
is to remove one little arch: if there is no marker inside the 
contribution is zero; otherwise,
substract one to the value $k$ of the marker inside, possibly removing it if it becomes zero,
and multiply the contribution by the corresponding $U_{k-1}$.

\example For $\Psi_{1,3,4,5}$, the non-zero coefficients are:
$$\figbox{4cm}{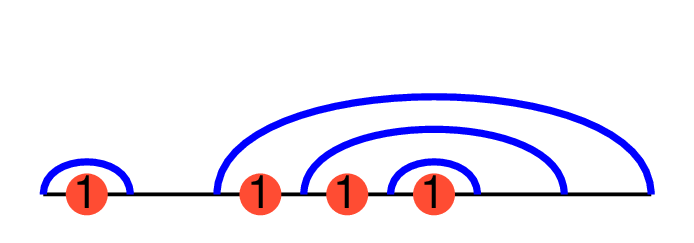}=\figbox{4cm}{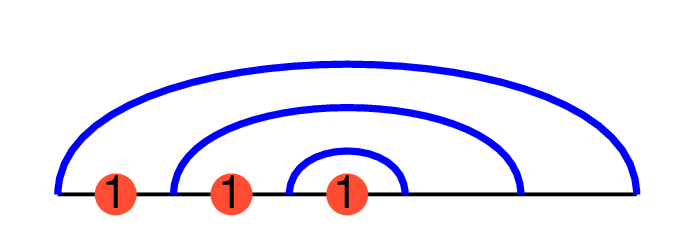}=\cdots=1
$$
$$\figbox{4cm}{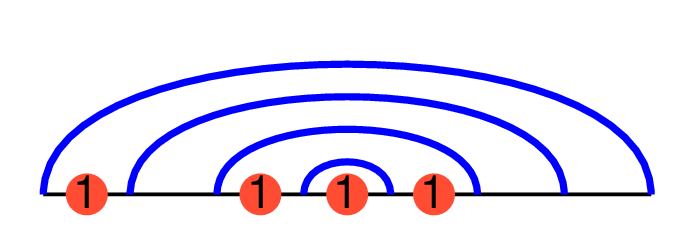}=\figbox{4cm}{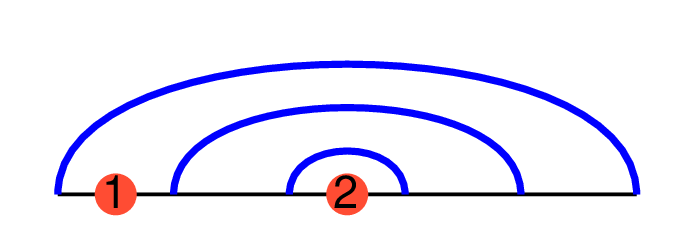}=U_1\figbox{4cm}{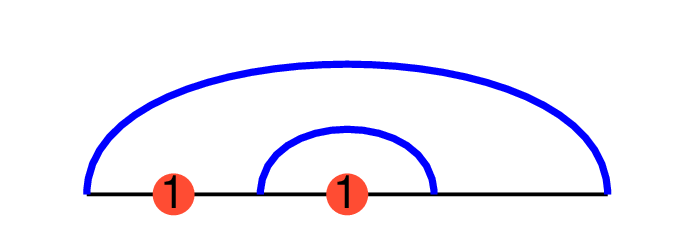}=\cdots=U_1
$$
$$\figbox{4cm}{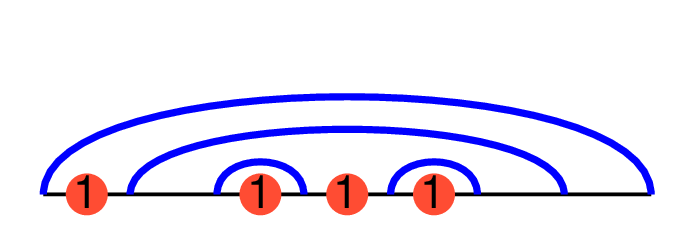}=\figbox{4cm}{ex2b.eps}=\cdots=1
$$

For $\Psi_{2,4,5,5}$:
$$\figbox{3.2cm}{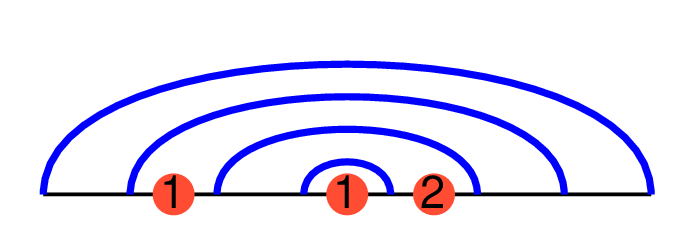}=\figbox{3.2cm}{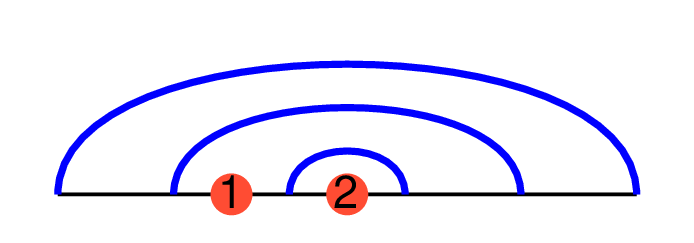}
=U_1\figbox{3.2cm}{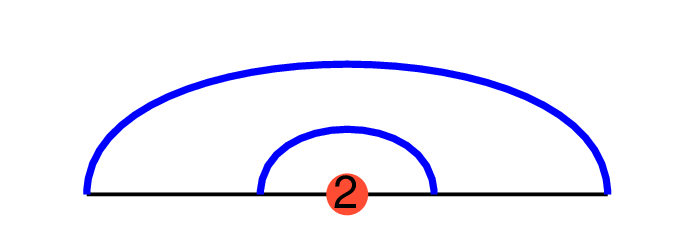}=U_1^2\hskip-0.5cm\figbox{3.2cm}{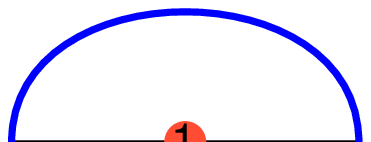}\hskip-0.5cm=U_1^2$$
$$\figbox{3.2cm}{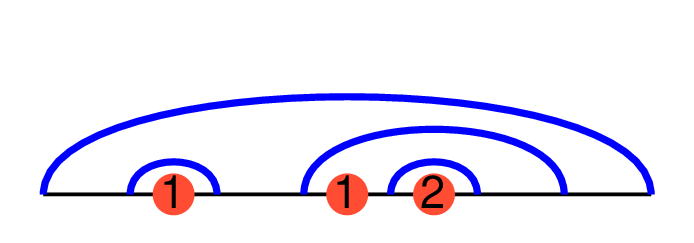}=\figbox{3.2cm}{ex4b.eps}
=U_1\figbox{3.2cm}{ex4c.eps}=U_1^2\hskip-0.5cm\figbox{3.2cm}{ex4d.eps}\hskip-0.5cm=U_1^2$$

\bigskip

The recurrence can of course be solved, and the final formula for the 
coefficient of $\Psi_\pi$ in $\Psi_{a_1,\ldots,a_n}$
is:
$$\prod_{i<\pi(i)} U_{\# \{ \ell: i\le a_\ell<\pi(i)\} - (\pi(i)-i+1)/2}$$
where $U_{-1}=0$, $U_0=1$, $U_1=-\tau$, $U_2=\tau^2-1$, etc.

As a check, note that this allows to recover Eq.~\partsumeq. Indeed, if $a_\ell\in\{2\ell-2,
2\ell-1\}$, one has
$$
\# \{ \ell: i\le a_\ell<\pi(i)\}=
\cases{ (\pi(i)-i+1)/2 & $i$ even and $\pi(i)-1\in \{a_\ell\}$
or $i$ odd and $i\in \{ a_\ell\}$\cr
(\pi(i)-i-1)/2&otherwise.\cr}
$$

\appendix{B}{Spin basis components}
The expression of the
solution of level $1$ $q$KZ equation in the spin basis (basis of sequences 
of $n$ $+$'s and $n$ $-$'s), as given in
\JM, once rid of various prefactors which are irrelevant for our purposes,
reads:
$$\eqalign{
&\tilde\Psi_{a_1,\ldots,a_n}=(q-q^{-1})^n
\prod_{1\le i<j\le N}(q\,z_i-q^{-1}z_j)\cr
&\oint\cdots\oint \prod_{\ell=1}^n {w_\ell\/ \d w_\ell\over 2\pi i}
{
\prod_{1\le \ell<m\le n} (w_m-w_\ell)(q\,w_\ell-q^{-1}w_m)
\over
\prod_{\ell=1}^n \prod_{1\le i\le a_\ell} (w_\ell-z_i) 
\prod_{a_\ell\le i\le N} (q\,w_\ell-q^{-1}z_i)
}}$$
in which the $a_\ell$ are the locations of the $+$'s. 
Noting that
$(q-q^{-1})w_\ell=(q\,w_\ell-q^{-1}z_i)-q^{-1}(w_\ell-z_i)$
and comparing with Eq.~\defpsi, we find:
\eqn\connec{
\tilde\Psi_{a_1,\ldots,a_n}=
\sum_{\varepsilon_1,\ldots,\varepsilon_n\in\{0,1\}}
(-q)^{-\sum_i \varepsilon_i}
\Psi_{a_1-\varepsilon_1,\ldots,a_n-\varepsilon_n}
}
In particular in the homogeneous limit we have the integral formula,
analogous to \coroleq:
$$\tilde\Psi_{a_1,\ldots,a_n}=
\oint\cdots\oint \prod_{\ell=1}^n {\d u_\ell(1-q^{-1}u_\ell)\over 2\pi i
\, u_\ell^{a_\ell}}
\prod_{1\le \ell<m\le n} (u_m-u_\ell)(1+u_\ell u_m+\tau u_m)
$$
Note that the formula for $a_\ell=2\ell-1$ (``largest component'' in 
the spin basis $+-\cdots+-$) is the special case $t=-q^{-1}$ of the
``refined enumeration'' of Eq.~\intc.

It is perhaps interesting that one can recover the change of basis
from link patterns to spins in our formalism. Indeed, using the
method of appendix A we can content ourselves with examining the
contribution of a little arch $\{ i,i+1\}$. The meaning of
Eq.~\connec\ is that each $+$ can be moved either one half-step to the right
with weight $1$, or one half-step to the left with weight $-q^{-1}$.
We have four local configurations around $i,i+1$:
$$\figbox{3cm}{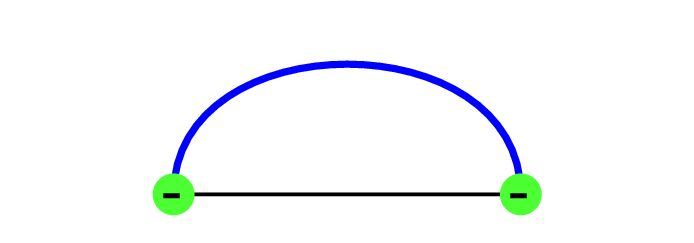}=\figbox{3cm}{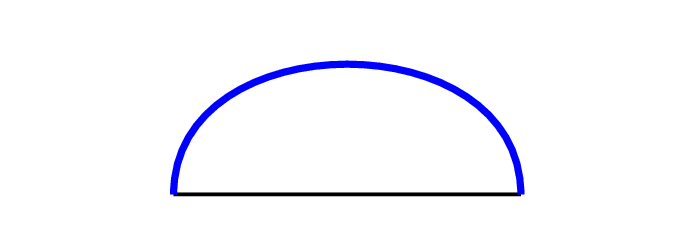}=0$$
$$\figbox{3cm}{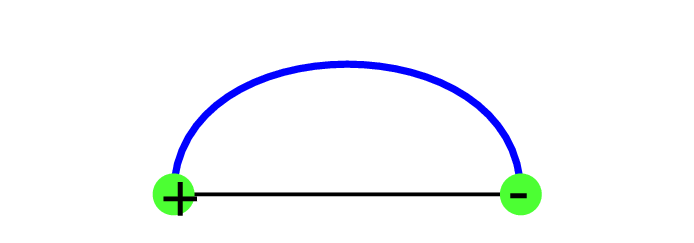}=\figbox{3cm}{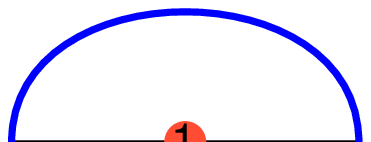}-q^{-1}\figbox{4cm}{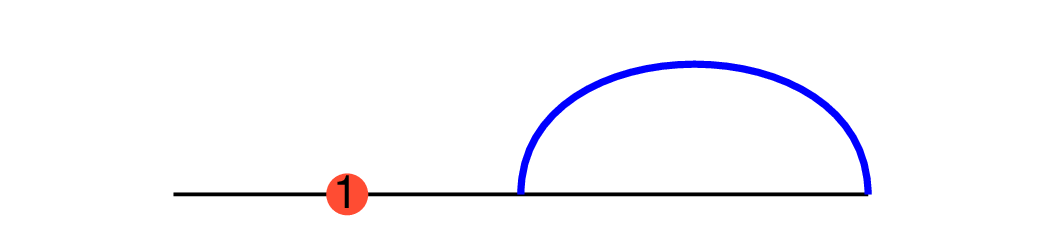}=1$$
$$\figbox{3cm}{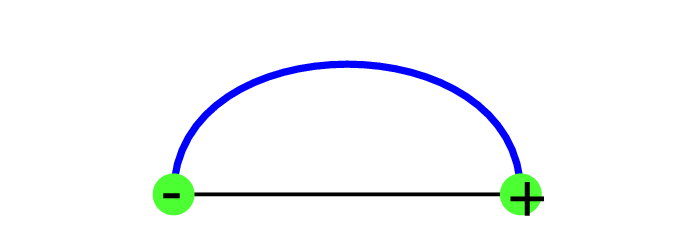}=\figbox{4cm}{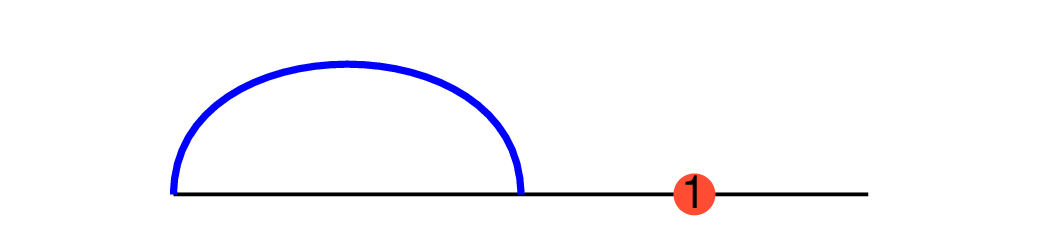}-q^{-1}\figbox{3cm}{spin2b.eps}=-q^{-1}$$
$$\figbox{3cm}{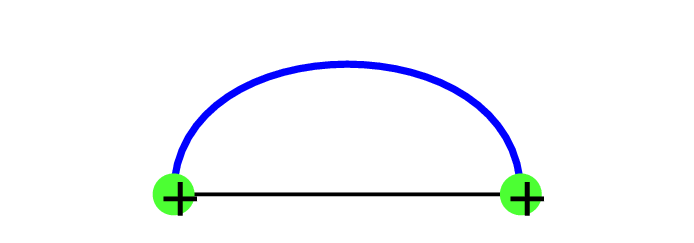}=\cases{
\figbox{4cm}{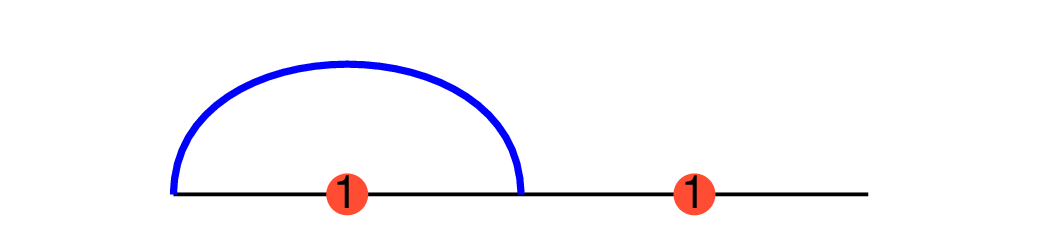}=\figbox{3cm}{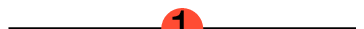}\cr
-q^{-1}\figbox{3cm}{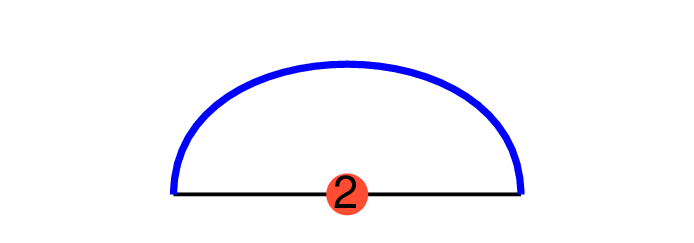}=-(1+q^{-2})\figbox{3cm}{spin4f.eps}\cr
-q^{-1}\figbox{5cm}{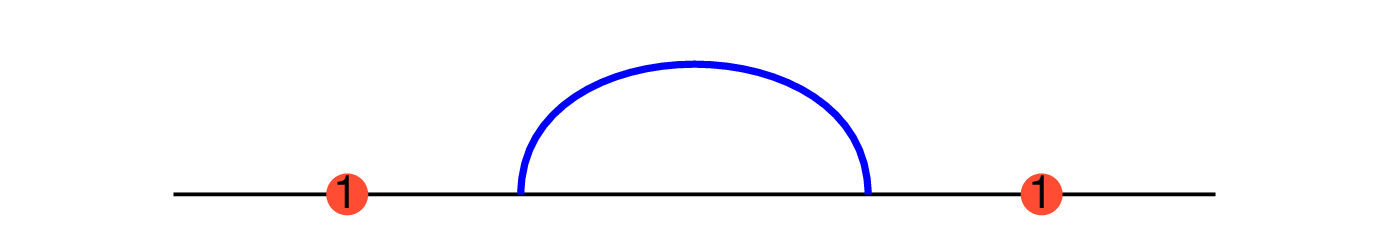}=0\cr
+q^{-2}\figbox{4cm}{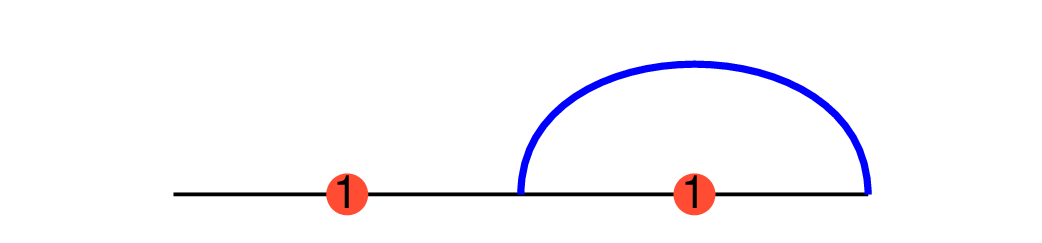}=q^{-2}\figbox{3cm}{spin4f.eps}\cr
}$$
and note that in the last case all contributions cancel, so that we
recover the usual rule that an arch gets a weight of $1$ if its opening
has a $+$ and its closing a $-$, a weight of $-q^{-1}$ if its opening
has a $-$ and its closing a $+$, and zero otherwise.

\end